\documentclass[reprint,superscriptaddress,amsmath,amssymb,aps,nofootinbib,prx]{revtex4-1}

\usepackage{amsthm}
\usepackage[colorlinks=true,urlcolor=blue,citecolor=blue,linkcolor=blue]{hyperref}
\usepackage[shortlabels]{enumitem}
\usepackage{amsfonts}
\usepackage{graphicx}
\usepackage[caption = false]{subfig}
\usepackage{multirow}
\usepackage{float}
\usepackage{bbold}
\usepackage{stmaryrd}
\usepackage{color}
\usepackage{ragged2e}
\usepackage{hyperref}
\usepackage{verbatim}
\usepackage{cases}
\usepackage{amsfonts}
\usepackage{amssymb}
\usepackage[]{mathrsfs}
\usepackage[table]{xcolor}
\usepackage{epsfig}
\usepackage{bm}
\usepackage{setspace}
\usepackage{enumerate}
\usepackage{dcolumn}
\usepackage{bm}
\usepackage{enumitem}
\usepackage{caption} 
\usepackage{tikz}
\usepackage{multirow, array} 

\usepackage{graphicx}
\usepackage{dcolumn}
\usepackage{bm}
\usepackage{color}
\usepackage{amssymb}
\usepackage{lipsum}
\usepackage{amsmath}
\usepackage{graphicx}
\usepackage{array}
\usepackage{booktabs}
\usepackage[english]{babel}
\begin{document}


\title{Exploring the Multifractal Behavior of the Human Genome T2T-CHM13v2.0: Graphical Representations and
Cytogenetics}

\author{Yulián A. Álvarez-Ballesteros}%
\affiliation{Departamento de Física, Universidad Autónoma Metropolitana Unidad Iztapalapa, Av. San Rafael Atlixco 186, Leyes de Reforma 1ra Secc, Iztapalapa, ciudad de México 09340, Mexico.}

\author{Mario A. Quiroz-Juarez}%
\email{maqj@fata.unam.mx}
\affiliation{Centro de F\'{i}sica Aplicada y Tecnolog\'{i}a Avanzada, Universidad Nacional Aut\'onoma de M\'exico, Boulevard Juriquilla 3001, 76230 Quer\'{e}taro, M\'exico}

\author{José L. Del-Rio‑Correa}%
\affiliation{Departamento de Física, Universidad Autónoma Metropolitana Unidad Iztapalapa, Av. San Rafael Atlixco 186, Leyes de Reforma 1ra Secc, Iztapalapa, ciudad de México 09340, Mexico.}

\author{Adrian M. Escobar-Ruiz}%
\affiliation{Departamento de Física, Universidad Autónoma Metropolitana Unidad Iztapalapa, Av. San Rafael Atlixco 186, Leyes de Reforma 1ra Secc, Iztapalapa, ciudad de México 09340, Mexico.}



\begin{abstract}
In this work, we applied the Chaos Game Representation (CGR) to the complete human genomic sequence T2T-CHM13v2.0, analyzing the entire chromosome assembly and each chromosome separately, including mitochondrial DNA. Multifractal spectra were determined using two types of box-counting coverage, revealing slight variations across most chromosomes. While the geometric support remained consistent, distinct distributions were observed for each chromosome. Chromosomes 9 and Y exhibited the greatest differences in singularity (Hölder exponent), with minor variations in their fractal support. The CGR distributions generally demonstrated an approximate separation between coding and non-coding sections, as well as CpG or GpC islands. A base-by-base analysis of the fractal support of the CGR uncovered characteristic structural bands in chromosome sequences, which align with patterns identified in cytogenetic studies. Using the complete assembly as a reference, we compared two alternative representations: the Binary Genomic Representation (RGB) and the Markov Chain (MC) representation. Both methods tended toward the same fractal support but displayed differing distributions based on the assigned length parameter. Multifractal analysis highlighted quantitative differences between these representations: RGB aligned more closely with high-frequency components, while MC showed better correspondence with low frequencies. The optimal fit was achieved using MC for twelve-base chains, yielding an average percentage error of $2\%$ relative to the full genomic assembly.
\end{abstract}

\maketitle



\section{Introduction} 

Deoxyribonucleic acid (DNA) is the molecule that contains the essential information needed to perform an organism’s vital functions. It is known for its double-helix structure, held together by chemical bonds between four nucleotides or bases: adenine (A), cytosine (C), guanine (G), and thymine (T), forming a sequence of bases called the Genomic Sequence (GS). The order of these bases along the primary structure of DNA encodes biological information. DNA is located within the cell nucleus, and during cell division, it is structurally organized into chromosomes \citep{genome}. In humans, under normal conditions, there are 24 different chromosomes, numbered from 1 to 22 as somatic chromosomes, along with the sex chromosomes X and Y.

\begin{figure*}[t!]
    \centering
    \includegraphics[width=\linewidth]{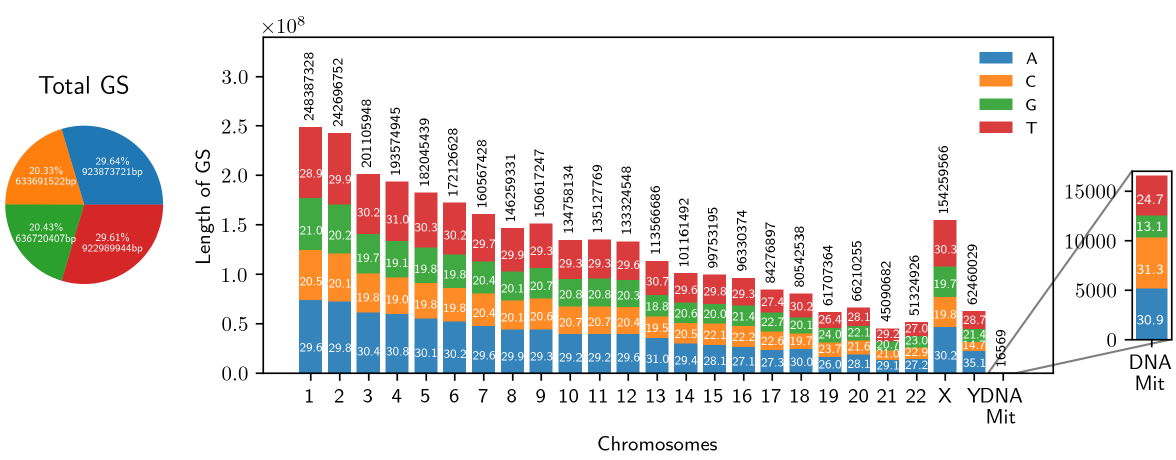}
    \caption{Approximate percentage of nucleotide bases per chromosome. Adenine (A) and thymine (T) show a higher prevalence compared to cytosine (C) and guanine (G).}
    \label{Fig:figure1}
\end{figure*}

Each chromosome has a unique structure that allows for differentiation, and the study of their number and structure is called cytogenetics (see Section \ref{sec: Citogenetica}), this is essential for diagnosing potential genetic disorders. Chromosomes have different lengths (see Figure \ref{Fig:figure1}), contain diverse genes and structures, and the information in each is responsible for specific functions, which is why decoding the genome is so important. These general characteristics create various patterns or base distributions in the GS, which we will explore from a multifractal approach in subsequent sections.

The T2T-CHM13 genomic assembly \citep{SG} is the first complete human GS; it corresponds primarily to female cells, with information for the first 22 chromosomes, the X chromosome, and mitochondrial DNA. The GS of the Y chromosome was taken from a male sample, thus completing the 24 chromosomes with a total of $3,117,275,501$ base pairs (bp), which occupies approximately $3.1 \, \text{Gb}$ of data.

Although the fractal or self-similar structure of the GS has been known for some time \citep{Jeffrey,Deschavanne,garte2004fractal,doi:10.1142/S0129065722500289,moreno2011human,ALBRECHTBUEHLER201220,cattani2013fractal,yu2001measure}, the lack of complete sequences without gaps prevented the construction of comprehensive representations for GSs with a large number of bases, which are needed to confirm this feature throughout the entire sequence and for each chromosome. In this work, we present graphical representations of the complete human genomic sequence and analyze two encoding methods: the binary genomic representation and the Markov chain representation. Using chaos game representation, we examined the multifractal properties of the full chromosome assembly and individual chromosomes, including mitochondrial DNA, uncovering distinct distributions and characteristic structural patterns. Chromosomes 9 and Y showed the greatest differences in singularity, while CGR distributions highlighted approximate separations between coding, non-coding regions, and CpG/GpC islands. Comparisons with RGB and MC revealed that RGB aligned with high-frequency components, whereas MC better captured low frequencies, achieving a $2\%$ error margin with twelve-base chains.

\subsection{Human Cytogenetics} 
\label{sec: Citogenetica}

\begin{figure}[t!]
    \centering
    \includegraphics[width=\linewidth]{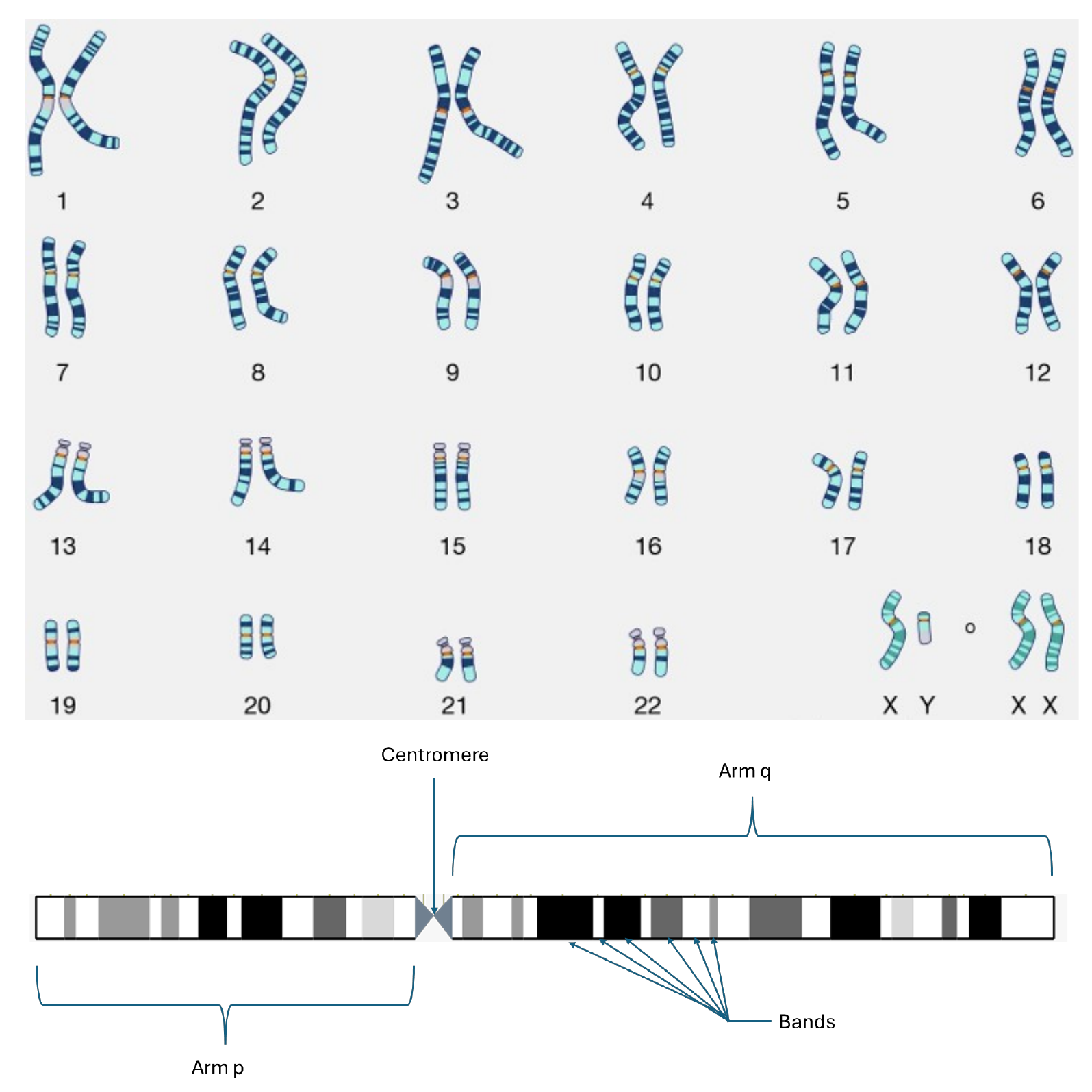}
    \caption{Karyotype image of an individual, displaying all human chromosomes and their structure, composed of four basic parts: the arm p, centromere, arm q, and the bands or stripes along the entire sequence. Source: "Courtesy: National Human Genome Research Institute". \citep{genome}}
    \label{Fig:figure2}
\end{figure}

The number and morphology of chromosomes in a cell of a specific species are generally constant in most cells of that organism (with the exception of reproductive cells and certain other cells, like those in the liver). This is a distinguishing characteristic of each species. Chromosome characterization is typically performed through chromosome banding techniques, which produce visible bands on chromosomes. These bands create a unique pattern for each chromosome (see Figure \ref{Fig:figure2}). This information can also be captured along the Genomic Sequence (GS) for each chromosome, and there are specialized databases where scientists collect and store new findings based on the GS.

As mentioned earlier, human GSs contain different types of functional and structural regions, each with a specific role in genome organization, regulation, and function. Below, we briefly summarize the main regions and their functions, information that allows us to approximate a function for each part of the graphic representations \citep{strachan2018}:

\begin{itemize}
    \item Genes and exons: Genes contain information necessary for producing proteins and functional RNAs. Exons are the parts of the gene that are transcribed and translated into proteins, while introns are removed during RNA processing.
    \item Promoters and regulatory elements: Promoters are sequences near genes that regulate the start of transcription. There are also enhancers, silencers, and other regulatory elements that modulate gene activity, influencing which genes are expressed and when. This means that enhancers increase the likelihood of gene expression, silencers reduce or prevent it, and other regulatory elements help determine the timing and conditions under which specific genes are activated or repressed, allowing cells to respond to different needs and environments.
    \item Non-coding regions: While these regions do not code for proteins, they are important for gene regulation. They include non-coding RNA sequences, like microRNAs and long non-coding RNAs (lncRNAs), which regulate gene expression and other cellular processes.
\end{itemize}

Repetitive regions: The human genome contains several repetitive sequences, including:

\begin{itemize}
    \item Microsatellites and minisatellites: Short, repetitive sequences that may vary between individuals.
    \item Interspersed repeats: Including SINE (Short Interspersed Nuclear Elements) and LINE (Long Interspersed Nuclear Elements), repeated sequences distributed throughout the genome that, in some cases, can move to other parts of the DNA.
    \item Centromeres and telomeres: Centromeres are specialized regions involved in chromosome separation during cell division. Telomeres are repetitive sequences at chromosome ends that protect the integrity of the genomic sequence, shortening with each cell division.
    \item CpG islands: Regions with a high density of cytosine and guanine dinucleotides, commonly found near gene promoters and associated with epigenetic regulation, such as methylation, which can silence gene expression.
\end{itemize}

In the context of genomic sequences, codons are sequences of three nucleotides that code for a specific amino acid in the translation process during protein synthesis. These triplets are found in the coding sequence of a gene (usually in exons) and are read by the ribosome in sets of three nucleotides, ensuring that each codon corresponds to an amino acid in the final protein.

There are 64 possible codon combinations: 61 of them code for specific amino acids, like methionine (ATG), which also acts as the start codon for translation, and 3 stop codons (TAA, TAG, TGA),  These codons signal the end of the protein sequence, instructing the ribosome to halt translation. In Figure \ref{Fig:figure4}, we show the location of the CGR points by region, which we will discuss in detail later. This distribution also allows us to locate codons with their corresponding amino acids.

The role of codons is essential, as they set the reading frame of the genetic sequence and ensure that the protein is correctly assembled according to the genetic information stored in DNA.

The aim of this article is to present an analysis of the human genome (GS) based on its fractal and multifractal structure, as observed in graphical representations described in Section \ref{Sec: GR}. This introduces, as a first challenge, the generation of these representations to verify behavior across extremely long sequences, such as those present in the human genome. Given the complexity of representing these sequences through points (Section \ref{sec: CGR}), a box-based or grid-based distribution approach is proposed, enabling the visual identification of fractal patterns.

The second objective of this work is to quantify and identify the behavior of different genomic regions in the CHM13v2.0 assembly using multifractal analysis techniques \label{sec: AM}. Additionally, the study aims to demonstrate that these representations exhibit a stripe-like structure, similar to that observed in chromosomes (Section \ref{sec: Citogenetica}).

\begin{figure}[t!]
    \centering
    \includegraphics[width=\linewidth]{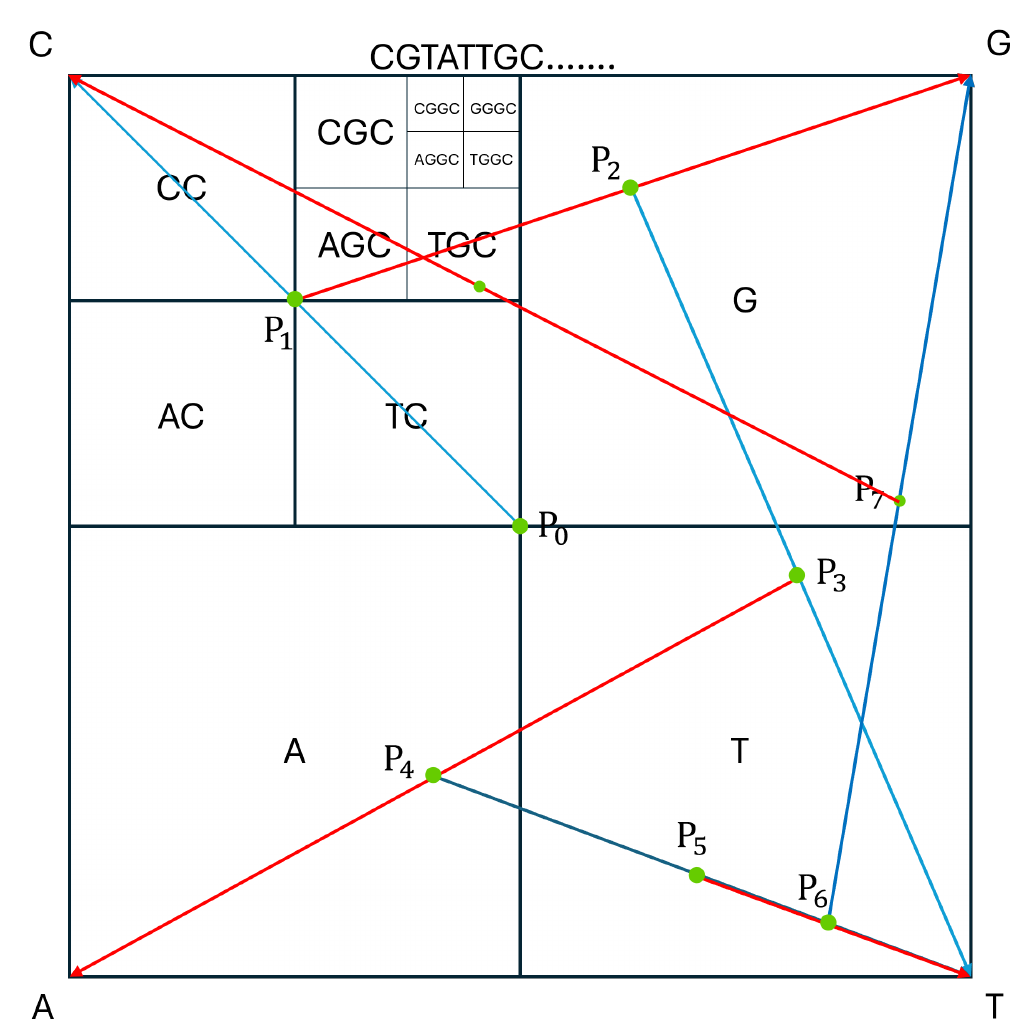}
    \caption{Example of applying the chaos game to an GS and dividing it by regions. In this case, we show in green the points corresponding to each base of the sequence that starts with the bases CGTATTGC. Point \(P_{1}\) is located halfway between the initial point \(P_{0}= (0.5, 0.5)\) and the corner corresponding to base C \(=(0, 1)\). Point \(P_{2}\) is located halfway between the previous point \(P_{1}= (0.25, 0.75)\) and the corner corresponding to base G \(=(1, 1)\), and so on for the other bases. This representation also creates a division into regions determined by the last bases of the sequence, so that any sequence ending in C will be in the corresponding quadrant \((0,0.5)\times(0.5,1)\). This region can also be subdivided into four regions for sequences ending in AC, CC, CG, and CT, which in turn can be further divided according to the four bases.}
    \label{Fig:figure3}
\end{figure}

This article begins with a description of genomic representation techniques (CGR, MC, and BGR) in Section \ref{Sec: GR}, which are commonly used to graph genomic sequences. Furthermore, the main concepts of multifractal analysis are introduced, including the $\tau$ and $q$ exponents, the generalized dimension, and the multifractal spectrum. These methods are applied to the CHM13v2.0 assembly in Section \ref{sec: res} Results and Analysis, positioning them as valuable tools to expand our understanding of the geometric characteristics of the human genome and, potentially, other species.

\begin{figure}[t!]
    \centering
    \includegraphics[width=\linewidth]{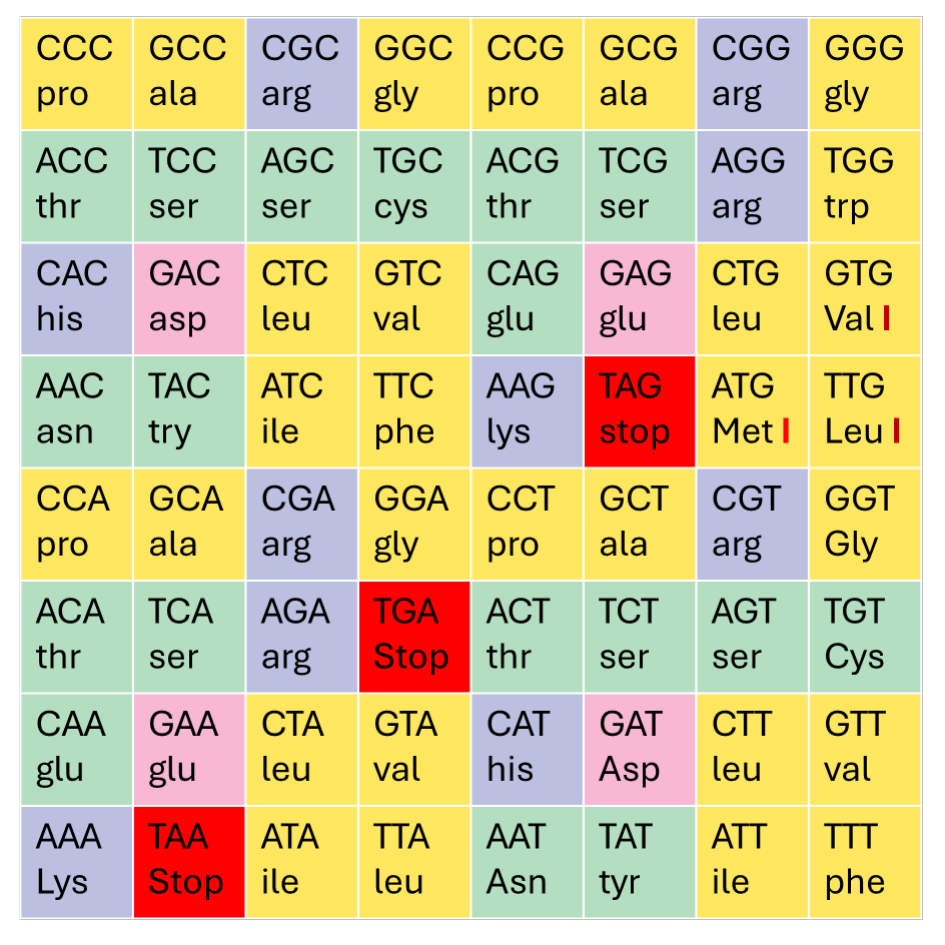}
    \caption{Correspondence zones between codons and amino acids (in symbols) in the CGR, including gene start and stop codons. The colors correspond to the four biochemical properties: nonpolar (yellow), polar (green), basic (purple), and acidic (pink). We observe that the C and T regions are symmetric with respect to the amino acids, meaning that these regions correspond to exactly the same amino acids, while the A and G regions show slight antisymmetry due to the presence of start and stop codons, maintaining a balance in the graphical representations.}
    \label{Fig:figure4}
\end{figure}

\subsection{Graphical Representations of Genomic Sequences}
The first method used to obtain images from genomic sequences (GS) \citep{Jeffrey} assigns a point within a square to each base of the GS, following the chaos game algorithm, which is equivalent to an iterated function system (IFS) \citep{Barnsley}. This approach allows for the generation of an image in the plane. These algorithms divide the square into smaller zones corresponding to specific short genomic sequences, facilitating the tracking of the GS within the image. The most important geometric characteristic is that the resulting structure is self-similar or fractal. This arises from the fact that the GS does not follow a homogeneous pattern in either the order or the quantity of different types of bases. 

\subsubsection{Chaos Game Representation}\label{sec: CGR}
The game of chaos is a mathematical visualization method used to show the principles of chaos theory \citep{strogatz2018nonlinear,escobar2022classical,quiroz2020experimental} and fractals \citep{feder2013fractals,barnsley2014fractals}. In our case, the method involves assigning the four bases (A, C, G, T) to the corners of the square. The midpoint is marked between the straight line segment from the center of the square to the corner corresponding to the first base of the GS. Then, the midpoint of the segment between the first point and the corner of the second base is marked, and this process continues until the entire GS is completed (see Figure \ref{Fig:figure3}).

\begin{figure}[t!]
    \centering
    \includegraphics[width=\linewidth]{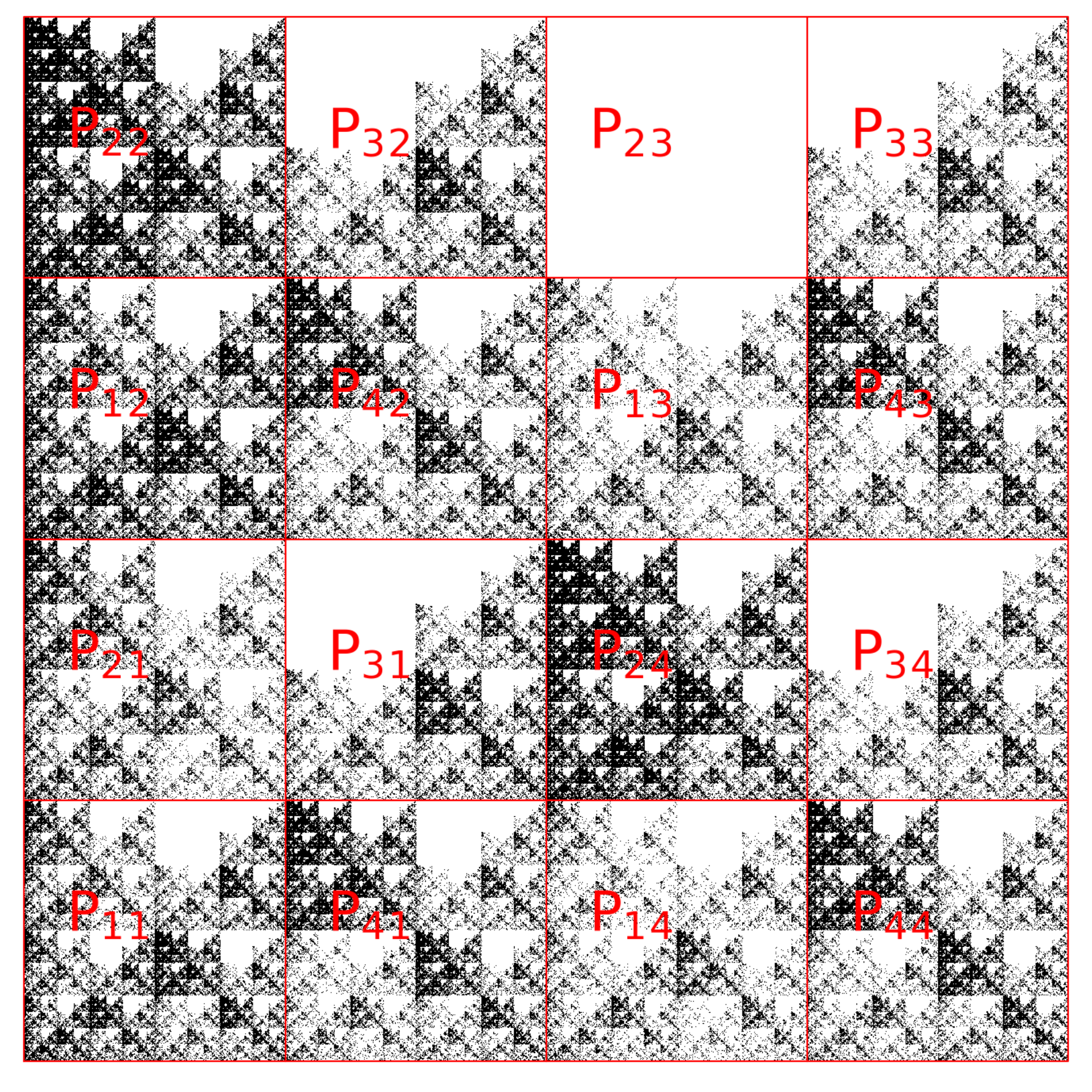}
    \caption{Example of point distribution in the CGR. In region $P_{23}$, the population is zero, which is reflected throughout the representation due to the absence of segments $P_{23}$.}
    \label{Fig:figure5}
\end{figure}

The pattern that can be obtained from different representations will depend on the distribution of the bases in the GS. Regardless of the length of the sequence, it is theoretically possible to trace in the CGR the origin and the sequence leading to each point, as in all representations, the space is divided into regions that can only be accessed through a given sequence (see Figure \ref{Fig:figure4}), meaning the GS can be represented by a single number \citep{Yin}. The computational limitation we face is that we cannot store infinitely large or small numbers, making it more convenient to continue using the original GS or the CGR.    

The low population or absence of a short sequence within the GS results in regions in the representation that are not visited \citep{Barnsley}. These regions self-replicate throughout the CGR (see Figure \ref{Fig:figure5}), allowing for the visualization of subsequences that are less, more present, or absent in the GS. 

This representation presents challenges when producing images; depending on the number of points, the limitations in the size of the visualization area, and the size of the points, the image tends to become saturated for a given length of the GS, making it difficult to distinguish the type of pattern obtained (see Figure \ref{Fig:figure6}). In some cases, it is useful to use opaque points, but this does not always work, as if opacity is not controlled by regions, the image becomes saturated in areas of high concentration or disappears in others. For this reason, for this and other representations, we divide the square into 412412 equal-sized cells and plot the number of points per cell (see Figure \ref{Fig:figure8}).

CGRs occupy a significant amount of memory (approximately 48 GB of storage for the complete human genomic sequence), which makes storage and processing difficult for computers that do not have sufficient memory. Therefore, it is advantageous to have efficient compression methods that allow the analysis of very large GSs and also leverage the patterns present in the GS, leading to simpler forms of encoding.

\subsubsection{Markov Chains}
Here, we construct Markov chains by assigning probabilities of obtaining any of the bases conditioned on the previous state. These previous states are defined as \(X\), which are short genomic sequences of fixed length \(n-1\) for \(n= 1, 2, 3, \ldots\), and the final state \(Y\in\{A, C, G, T\}\). 

For example, for \(n=2\), \(XY\) is a dinucleotide, meaning a sequence of two bases, where \(X, Y \in \{A, C, G, T\}\). For \(n=3\), \(X \in \{AA, AC, AG, AT, CA, CC, CG,\ldots, TT\}\) corresponds to all possible states of two bases, always with \(Y\in\{A, C, G, T\}\), and similarly for other values of \(n\). The probability matrix of the corresponding Markov chain (MC) will depend on the length \(n\) and is given by \citep{Goldman}:

\begin{figure*}[t!] 
\centering 
\includegraphics[width=\linewidth]{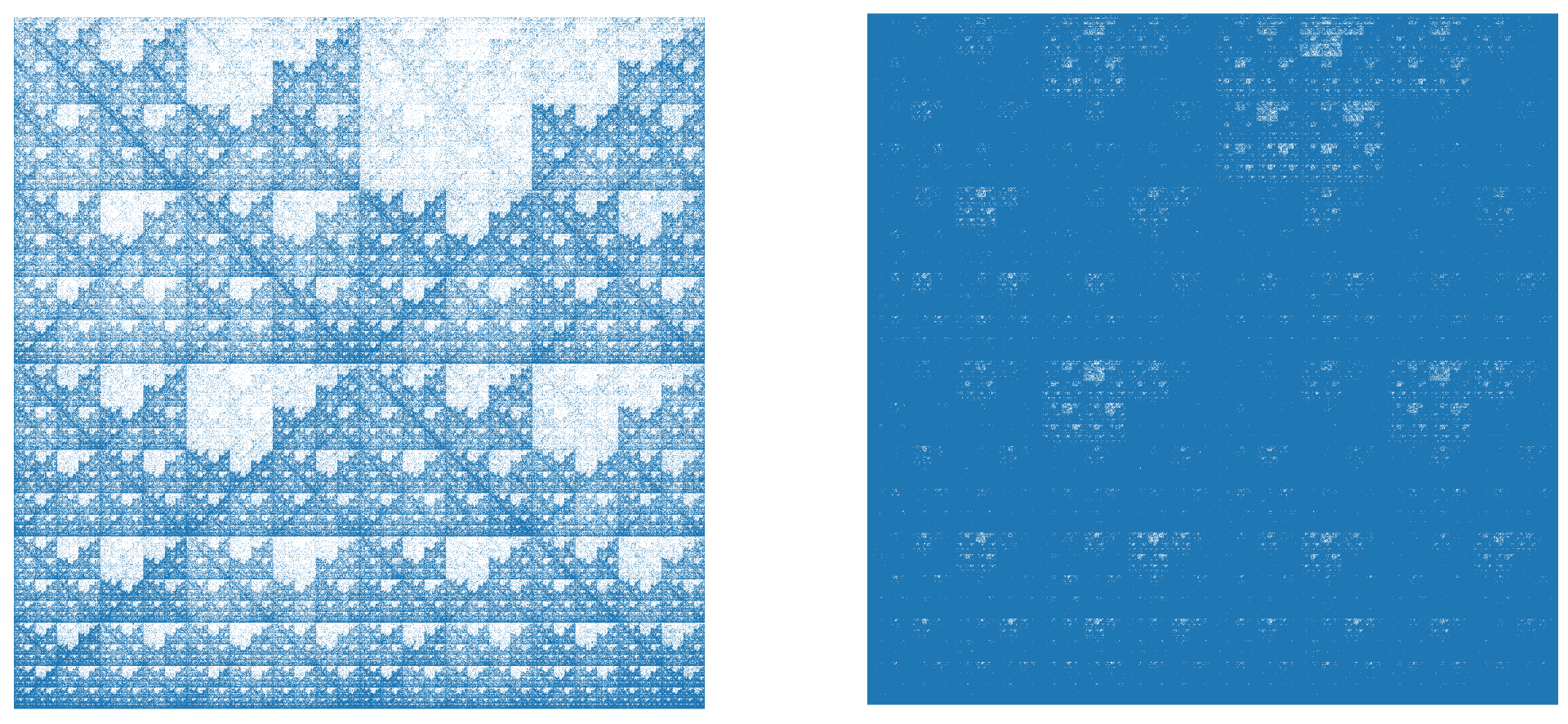}
\caption{On the left, an image of a CGR with a number of points on the order of $10^{6}$; on the right, on the order of $10^{9}$, where the pattern is no longer distinguishable.} 
\label{Fig:figure6} 
\end{figure*}

\begin{equation}
    P_{XY} = \frac{n_{XY}}{n_{XA}+n_{XC}+n_{XG}+n_{XT}},
\end{equation} 

where \(n_{XY}\) is the number of times \(XY\) appears in the GS. This process produces a probability matrix of size \(4^{n-1}\times 4\), generating a MC iteratively by adding a base at each step, dependent on the corresponding probabilities of the previous state. The new GS is then used to obtain a new CGR. Figure \ref{Fig:figure5} is an example of a CGR generated using an MC, where the probability \(P_{23}=0\).

\begin{figure}[t]
    \centering
    \includegraphics[width=\linewidth]{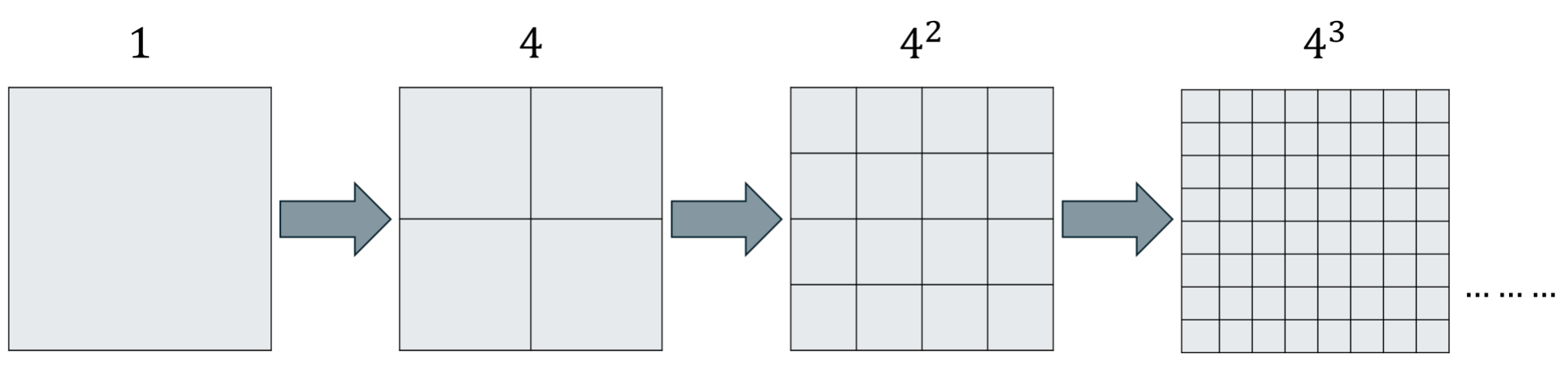}
    \caption{Division of the unit square used in the box-counting method: This division is achieved through the successive splitting of the square into vertical and horizontal halves. Each step produces a new set of disjoint squares of equal size, whose union reconstructs the original square. With each iteration, the total number of boxes quadruples.}
    \label{Fig:figure7}
\end{figure}

Since these are random systems originating from the GS, the MC does not allow for direct decoding back to the GS, which is its main disadvantage. However, depending on how well it fits the original distribution, a direct analysis can be performed on the probability matrix, which, depending on \(n\), may have significantly fewer elements than the GS. In this work, we will consider only four different values of \(n\) (\(2, 3, 6, 12\)), so the maximum number of elements in the probability matrices is \(4^{12}=16777216\).

Another aspect to consider in the MC is its length, ensuring it is sufficiently large to generate a distribution close to the original; in other words, how quickly we reach the attractor, assuming that the distribution of bases in the GS is situated on a fractal support.

\subsubsection{Binary Genomic Representation}

Inspired by the ideas proposed by \citep{Yin}, we propose a new method for compressing and representing GSs, consisting of assigning two-dimensional binary vectors to each of the bases:
\begin{equation}
    A= \begin{pmatrix} 0 \\ 0 \end{pmatrix}, \quad
    C= \begin{pmatrix} 0 \\ 1 \end{pmatrix}, \quad
    G= \begin{pmatrix} 1 \\ 1 \end{pmatrix}, \quad
    T= \begin{pmatrix} 1 \\ 0 \end{pmatrix}. 
\end{equation}

In this way, each point on the plane is represented in binary form, and the points in the BGR are defined as:
\begin{equation}
    P_{N} = 0.V_{N}V_{N-1}V_{N-2}\dots V_{2}V_{1}V_{0},
\end{equation}
where \(V_{j}\in(A, C, G, T)\). This point includes the previous point in the sequence, as:
\begin{equation}
    P_{N-1} = 0.V_{N-1}V_{N-2}\dots V_{2}V_{1}V_{0},
\end{equation}
Therefore, the points in the representation are obtained by shifting the binary point to the right. This way, BGR allows us to store the GS as two sequences of zeros and ones. It is clear that it is always possible to revert to the original GS.

The way the binary representation is encoded is through the segmentation of the GS; each of the segments of length \(M\) represents a point in the plane, so the GS is generally represented by \(n+1=\left\lfloor\frac{\#\text{ de bases en la GS}}{M}\right\rfloor+1\) points.

\begin{figure*}[t!]
    \centering
    \includegraphics[width=\linewidth]{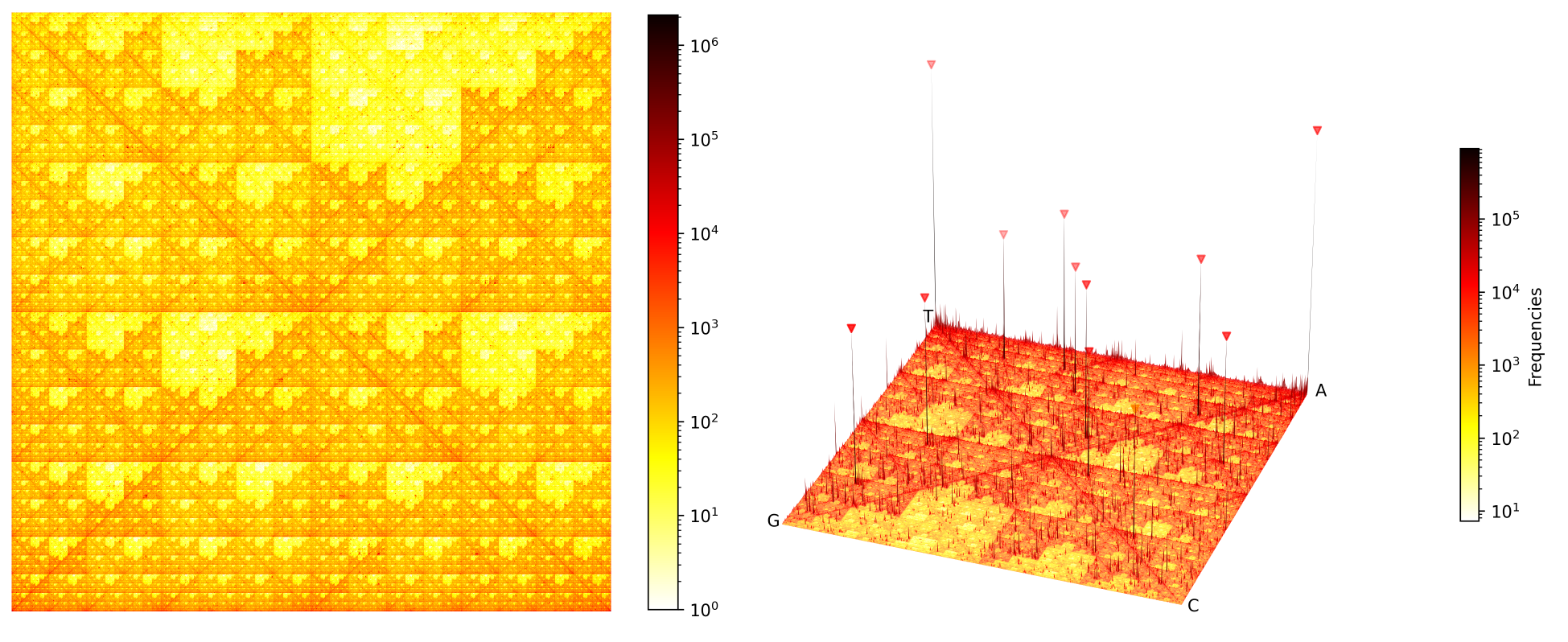}
    \caption{Logarithmic scale plot of the number of points per cell in the CGR, constructed using the methodology described in Section \ref{sec: CGR} for the CHM13v2.0 GS, in both 2D and 3D representations. The two-dimensional representation consists of a total of \(4^{12}\) cells, while the three-dimensional representation, designed to enhance the visualization of peaks, uses \(4^{10}\) cells.}
    \label{Fig:figure8}
\end{figure*}

\subsection{Multifractal Analysis}\label{sec: AM}

Different representations of the geometric series (GS) serve as fundamental tools for encoding and decoding, exhibiting qualitative characteristics such as self-similarity and distribution over geometric supports, which confer a multifractal nature to the GS \citep{harte2001}. Although there is no unique definition of fractality, it is understood that these sets can be assigned a non-integer dimension, which can be measured in various ways \citep{Barnsley}. In this study, we utilize the box-counting method, the same technique employed for randomly generated curves (CGR), making it quite convenient.

Since CGRs are represented in a plane, their dimensions have a maximum range of $[0, 2]$. However, due to the presence of empty spaces or regions within CGRs, their effective range may fall below this maximum. We can combine this dual structure of measures (frequencies or probabilities) and dimensions (fractal support) through the use of the multifractal spectrum $f(\alpha)$, where $\alpha$ is the Lipschitz-Hölder exponent \citep{harte2001}. The multifractal spectra associated with the CGRs is utilized to quantify differences between the ensemble, the chromosomes that compose it, and the results obtained from Monte Carlo (MC) and BGR methods.

The relationships used in this analysis, which allow us to obtain the spectrum, are based on two exponents associated with frequencies and geometric support: $q$ and $\tau$, respectively. These exponents are related by the following equation:
\begin{equation}\label{ec: tau} \tau(q) = \lim_{r\to 0}\frac{\log\sum_{l=1}^{N}p_{l}^{q}}{\log r}, \end{equation}
where $N$ is the number of covers, $r$ is the box size and $p_{l}$ is the number of visits to the $l$-th box relative to the total number of points in the representation. Another useful quantity for the analysis is the Rényi or generalized dimension \citep{schroeder}:
\begin{equation}\label{ec. dq} D_{q} = \frac{\tau}{q-1}, \end{equation}
which contains the same information as $\tau$ but assigns a fractal dimension to each value of $q$. Finally, the multifractal spectrum is obtained through the Legendre transform:
\begin{equation}\label{ec: sf} f(\alpha) = \inf_{q}{\left(q\alpha-\tau(q)\right)}, \end{equation}
where the following relationships are satisfied:
\begin{equation} \alpha = -\frac{d\tau}{dq},\quad q = \frac{df}{d\alpha}. \end{equation}

\begin{figure*}[t!]
    \centering
    \includegraphics[width=\linewidth]{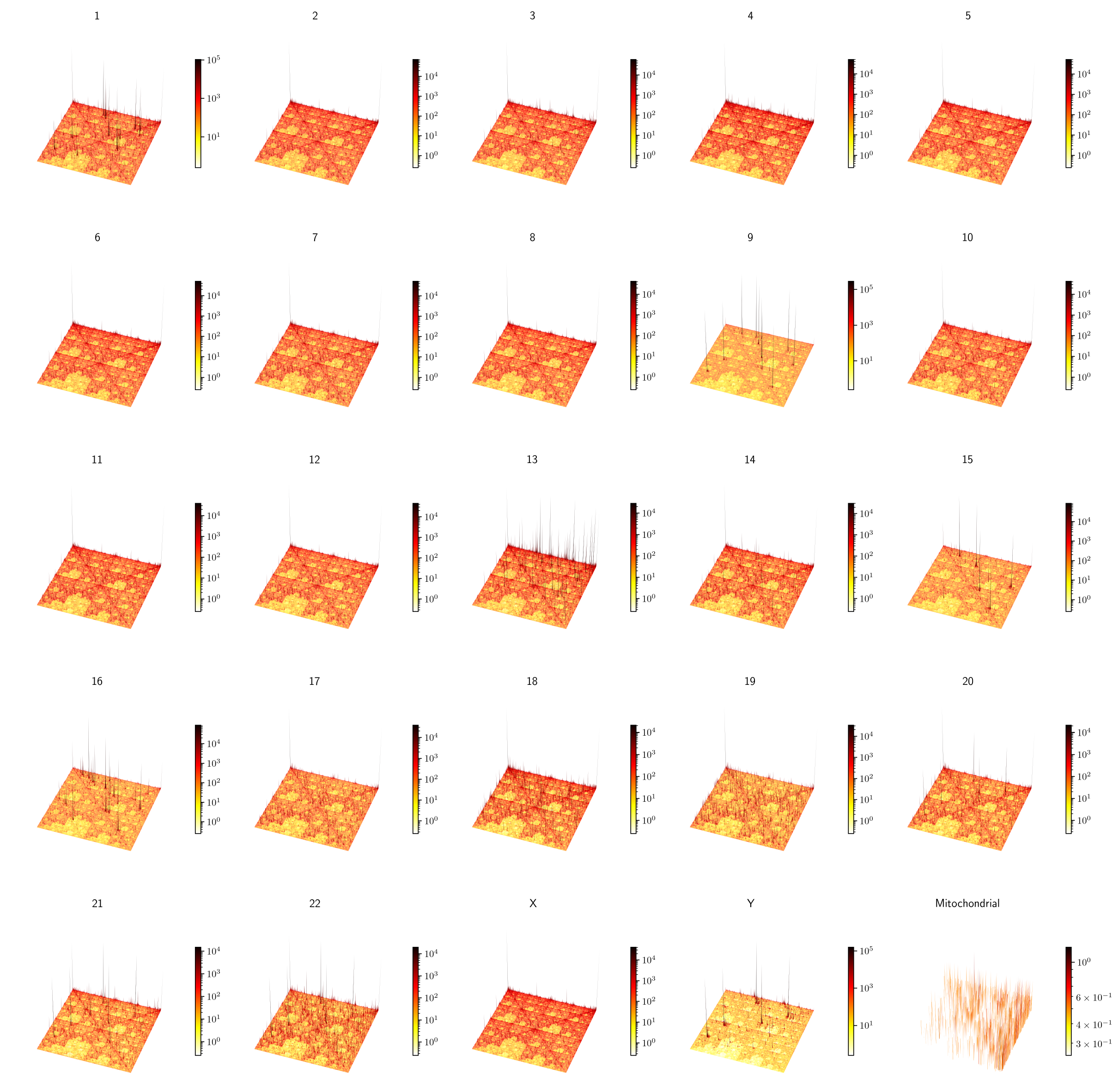}
    \caption{Graphs of the number of points per cell in the CGR for each chromosome and mitochondrial DNA.}
    \label{Fig:figure9}
\end{figure*}

\section{Methodology}

This study exclusively utilizes data from the GS CHM13v2.0 and the information provided in \citep{SG}. The methodology primarily focuses on generating the graphical representations outlined in Section \ref{Sec: GR}, which highlight the multifractal structure of the GS, both for the entire GS and its partition into the 24 chromosomes and mtDNA.

The analysis allows for the identification of point distributions or occupancy within the CGR, subdivided into cells. This procedure, known as box counting \citep{Barnsley}, involves recursively subdividing the square containing the CGR into smaller disjoint squares (see Figure \ref{Fig:figure7}). Each square is then assigned a value representing the number of CGR points contained within it. In this study, subdivisions of $4^{12}$ and $4^{10}$ boxes were applied to assess the point distributions in the CGRs of the GS.

These distributions were subsequently employed in the multifractal analysis as a means to compare and identify quantitative differences and similarities in the GS. For this purpose, the $p_{i}$ values were computed for each box, and $\tau$ was derived using equation \ref{ec: tau} for $q$ values ranging from $-30$ to $30$. Additionally, the fractal dimensions and multifractal spectrum were determined using Equations (\ref{ec. dq}) and (\ref{ec: sf}).

\begin{figure*}[t!]
    \centering
    \includegraphics[width=\linewidth]{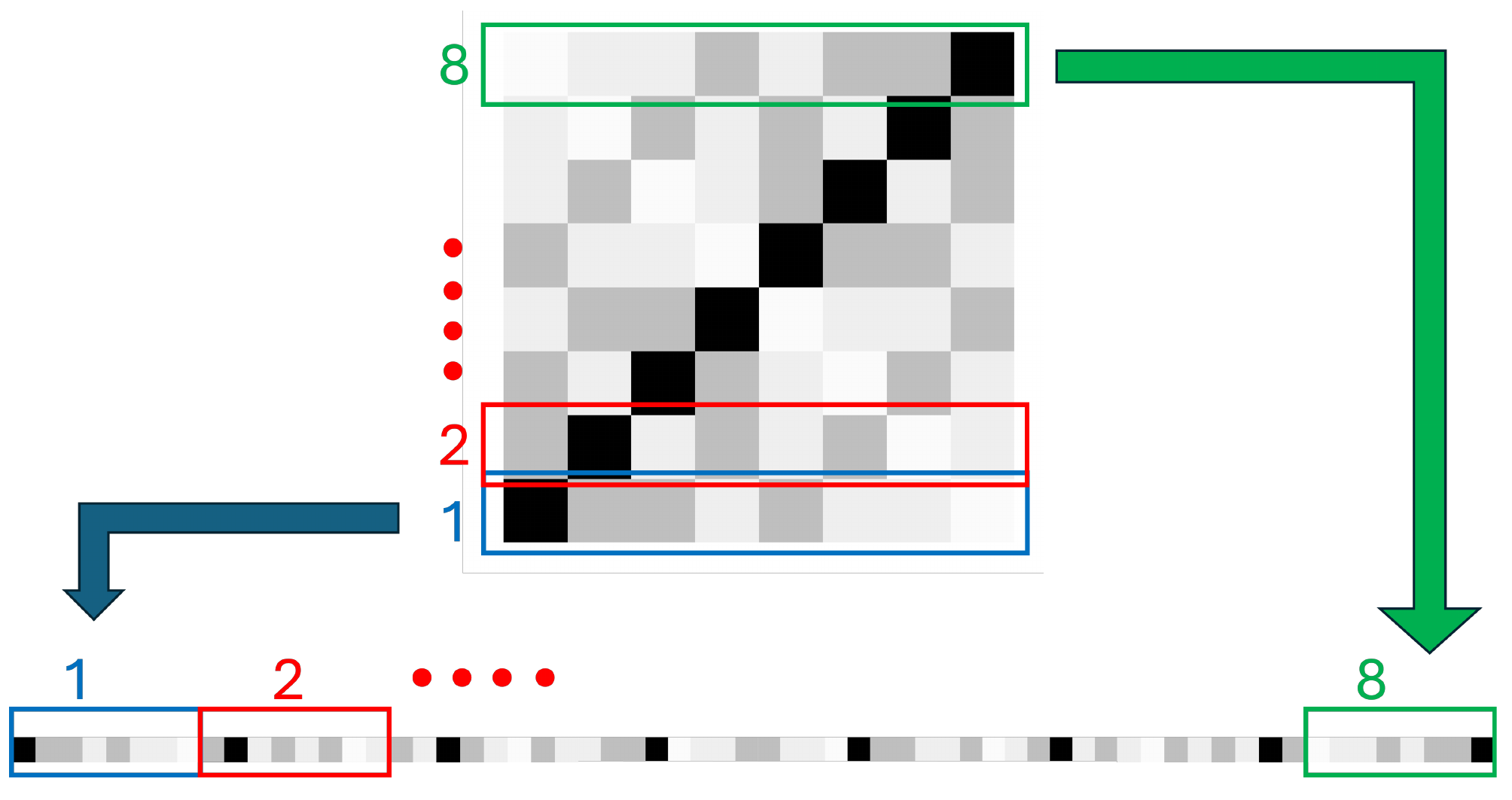}
    \caption{Example of converting a two-dimensional representation to a one-dimensional one. In this case, it shows the distribution of a CGR divided into $4^{3}$ square covers, where the color intensity in each quadrant reflects the corresponding measure in that region. For the conversion, we take each row, from the bottom to the top, and align them along a line. This procedure is routinely applied to CGRs obtained from the GS, which are divided into $4^{12}$ covers and $4^{6}$ rows.}
    \label{Fig:figure10}
\end{figure*}

We also generated one-dimensional CGRs, as shown in Figure \ref{Fig:figure10}, equivalent to the original representations. This approach allows us to gather the results obtained individually for the complete assembly, chromosomes, and mitochondrial DNA, enabling an approximate identification of differences in the distributions and the regions where they occur.

In Section \ref{sec: Citogenética de la CGR}, we describe the methodology used to derive an alternative one-dimensional visualization of the CGR, where each base is represented as part of a time series (see Figure \ref{Fig:figure15}).

\section{Results and Analysis}\label{sec: res}

Images of the (CGR) for the (GS) from the T2T-CHM13v2.0 assembly were obtained, along with those for individual chromosomes and mitochondrial DNA (see Figures \ref{Fig:figure8} and \ref{Fig:figure9}). To maintain the fractal structure and account for the predominance of certain regions, a logarithmic scale was employed. Although mitochondrial DNA is not part of the assembly and is not the main focus of this study, we included it to illustrate that its multifractal structure is distinctly different, exhibiting a higher presence of base C, which causes its support to resemble a Sierpinski triangle \citep{Barnsley}.

Since the length of the mitochondrial DNA sequence is significantly shorter than that of the chromosomes, it is appropriate to consider an analysis with fewer covers. Based on the distribution behavior concerning the number of boxes and computational constraints, we find that starting from $4^{10}$ provides a good approximation to the attractor of the distribution. 

Regarding the distribution of bases within the sequences, we observe a slightly higher number of A and T bases, although the difference is not disproportionate across any of the chromosomes. In terms of dinucleotides, the CGR clearly shows that the region corresponding to CG has the fewest points, and this absence is reflected throughout the entire CGR, particularly within itself. This feature suggests a significantly reduced probability of the base C being immediately followed by the base G. This understanding of sequencing will be more easily visualized through the MC.

The scarcity of CG dinucleotides in the representation has already been reported in other vertebrates \citep{Deschavanne}, along with the existence of CG or GC islands, suggesting that this representation may serve as a genomic signature for various species. Additionally, it is noteworthy that the significant peaks in the distribution, primarily composed of A and T bases in most chromosomes, generally correspond to long repetitive sequences within the genome, indicating a clear separation between coding and non-coding regions.

\begin{figure*}[t!]
    \centering
    \includegraphics[width=\linewidth]{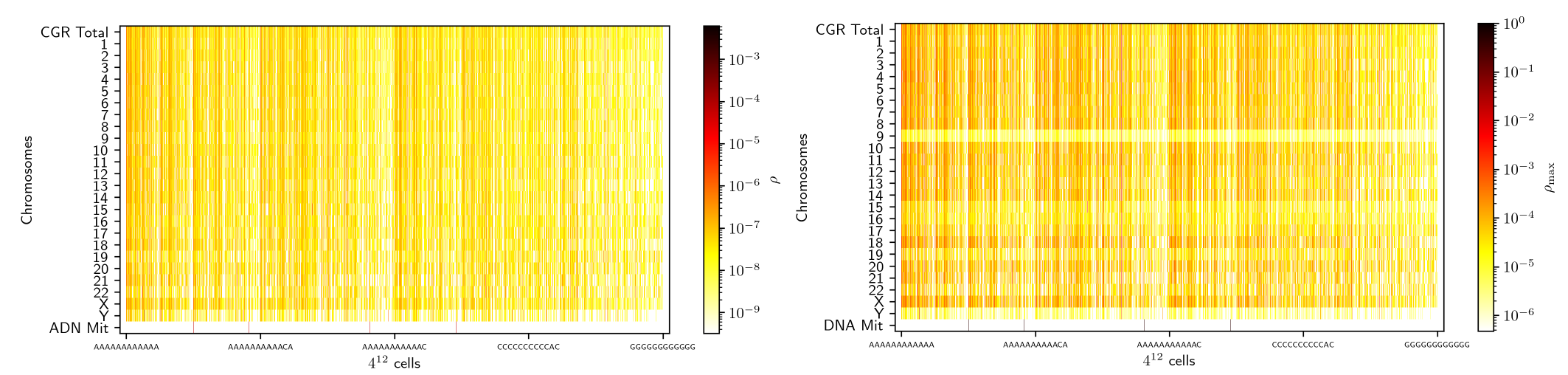}
    \caption{One-dimensional comparison of each chromosome with the complete assembly (Total CGR). The densities $\rho$ and $\rho_{\max}$ are shown on a logarithmic scale. We follow an ascending order in the CGRs, starting from the row corresponding to the short 12-base GS ranging from AAAAAAAAAAAA to TTTTTTTTTTTT, up to the row of CCCCCCCCCCCC to GGGGGGGGGGGG.
    }
    \label{Fig:figure11}
\end{figure*}

In Figure \ref{Fig:figure11}, we generated a one-dimensional 
arrangement for the CGRs of both the complete assembly and the individual chromosomes. We calculated the density $\rho$ (the number of points in each cell relative to the length of the respective genomic sequence) and the maximum density $\rho_{\max}$ (the number of points in each cell compared to the cell with the highest point count in the corresponding CGR). This normalization process adjusts all representations to unit maxima. This visualization also clearly demonstrates that the total CGR is essentially a superposition of the individual chromosomal CGRs.

\begin{figure*}[t!]
    \centering
    \includegraphics[width=15cm]{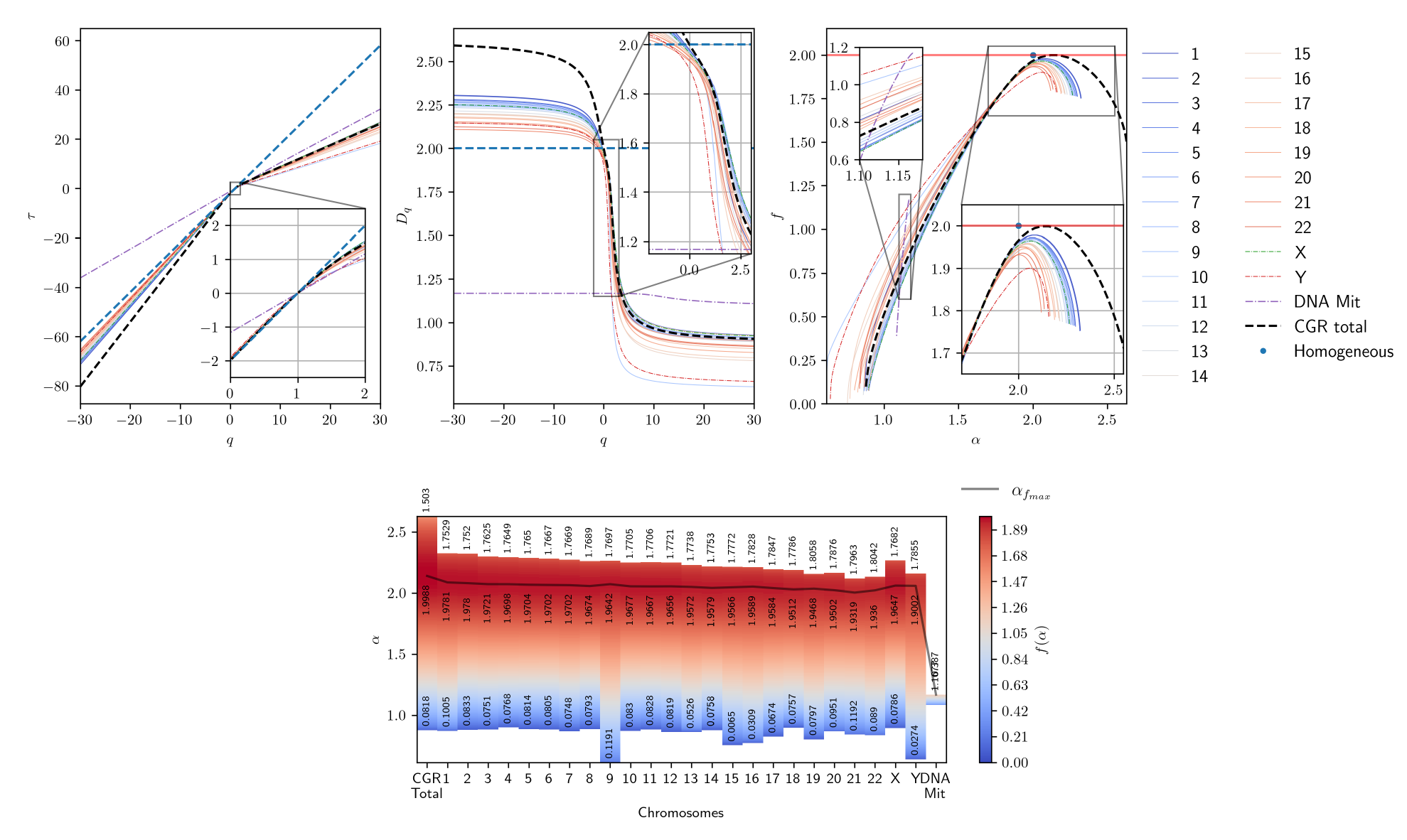}
    \caption{Multifractal analysis: the exponent \(\tau\) as a function of \(q\), the generalized dimension \(D_q\), and the multifractal spectrum \(f(\alpha)\), where \(\alpha\) represents the singularity exponent as defined in Section \ref{sec: AM}. The plots for each chromosome are presented and compared to the complete assembly. The behavior across different chromosomes is generally similar, with the exception of the Y chromosome and mitochondrial DNA.}
    \label{Fig:figure12}
\end{figure*}

In Figure \ref{Fig:figure12}, the graphs of the exponent $\tau$, the generalized dimension, and the multifractal spectrum are shown. Although all of them contain almost the same information, we decided to present them together with the intention of achieving greater distinction between the different results. However, we will focus on the multifractal spectrum, for which we provide a summary of each spectrum.

\begin{figure*}[t!]
    \centering
    \includegraphics[width=15cm]{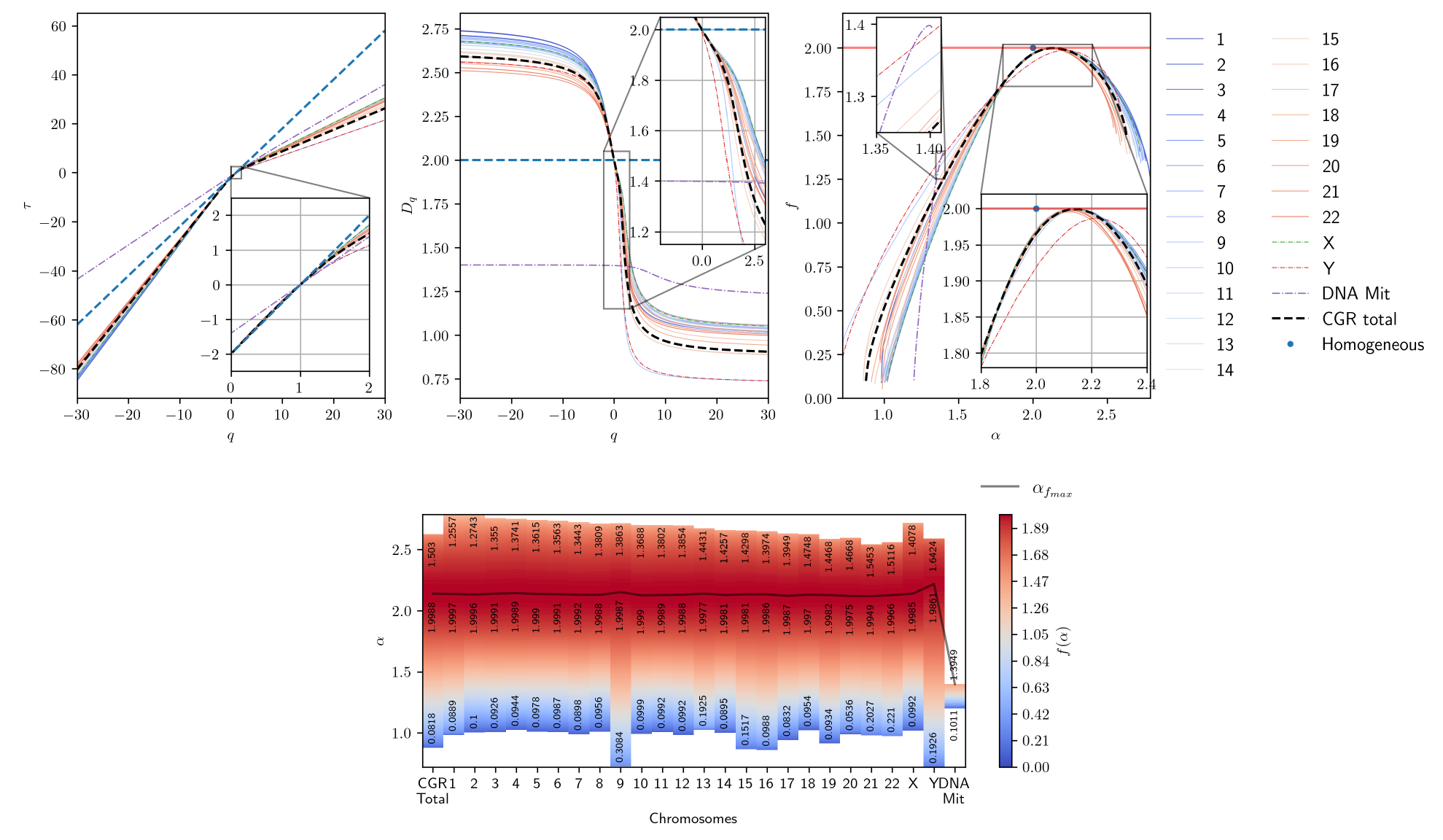}
    \caption{Multifractal analysis by chromosome compared to the complete assembly (in the same manner as shown in Figure \ref{Fig:figure12}). The number of coverings for the chromosomes was reduced to \(4^{10}\) to improve the fit with the assembly for each curve.}
    \label{Fig:figure13}
\end{figure*}

Clearly, the measures are significantly affected because the length of the GS corresponding to the assembly is much greater than the lengths of the chromosomes. To approximately correct this disparity, we considered a smaller number of covers for the chromosomal GSs. Figure \ref{Fig:figure13} shows the results, where a better fit can be observed in the area around $\alpha=2$, which is the area of greatest interest, albeit with losses at the extremes.

\begin{figure*}[t!]
    \centering
    \includegraphics[width=\linewidth]{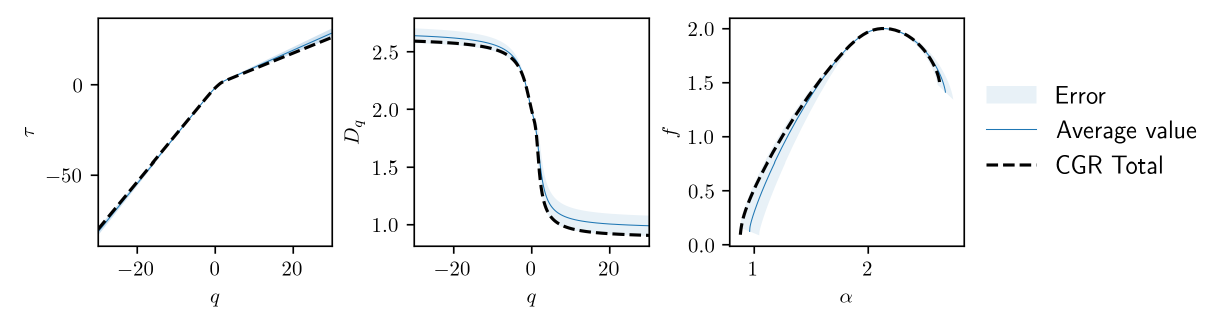}
    \caption{Comparison between the CGR assembly and the chromosomal average obtained from the results in Figure \ref{Fig:figure13}.}
    \label{Fig:figure14}
\end{figure*}

\begin{figure*}[t!] 
    \centering 
    \includegraphics[width=\linewidth]{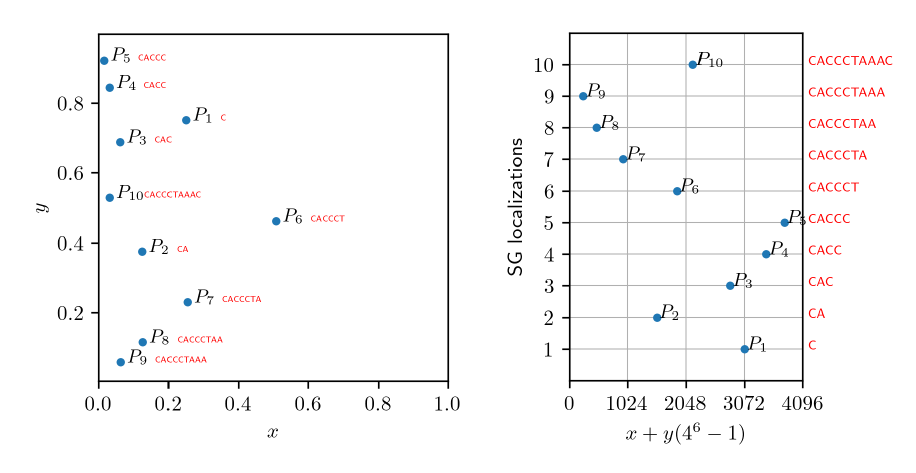}
\caption{On the left, the plot of the initial points of the GS for chromosome 1 using CGR, and on the right, the plot of the GS base location along the y-axis compared to the one-dimensional CGR obtained using the transformation \(x + \lfloor y(4^{6} - 1)\rfloor\) along the x-axis.} 
\label{Fig:figure15} 
\end{figure*}

From the results, we can highlight that although the behavior in the chromosomes tends to be the same, there is a clear bias in chromosomes 9 and Y, which are distinguished by having the greatest widths within the spectrum, not associated with the length of their GSs. Additionally, chromosome Y has the smallest maximum, which is slightly shifted to the right of the other spectral maxima. The average spectrum between chromosomes and its deviation was obtained (see Figure \ref{Fig:figure14}). The assembly remains within the deviation along the spectrum, with better adjustment in the area around the maximum.

The maxima in the spectra approach the dimension of the plane; however, since it is a finite set of points, in the infinitesimal limit, the spectra tend to have a dimension of zero. Therefore, the multifractal spectrum serves as an approximation, assuming the sequence is considerably large relative to the number of covers and sufficiently small to approach the attractor of the distribution. Nonetheless, given that the ensemble of chromosomes is significantly larger than individual chromosome sequences, we might assume that its average behavior would tend to remain consistent, even if chromosome lengths varied.

\begin{figure*}[t!] 
    \centering 
    \includegraphics[width=17cm]{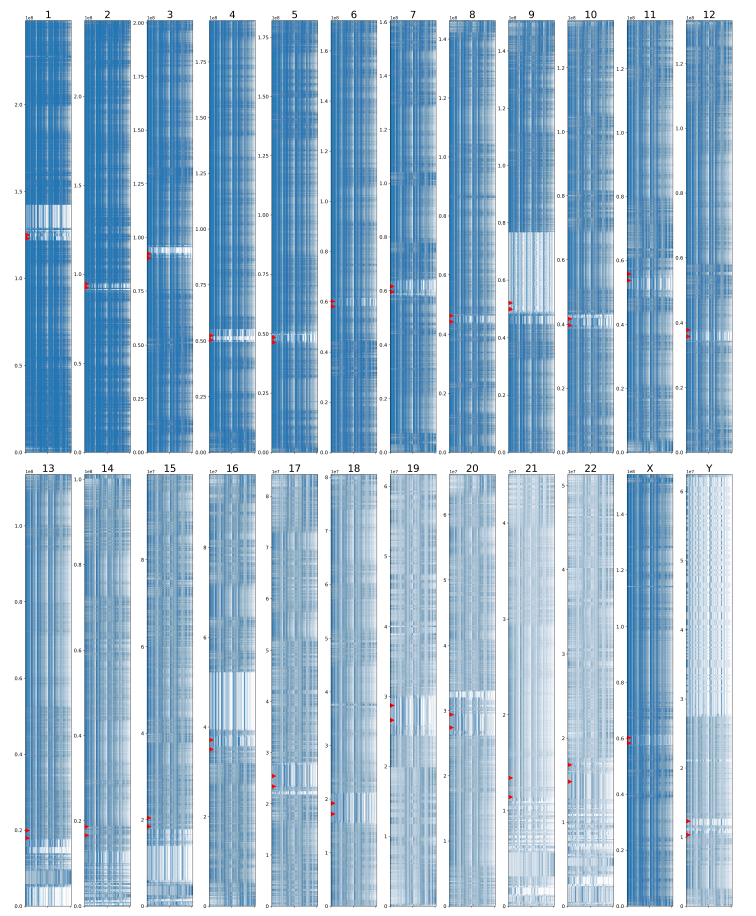}
    \caption{Complete GS for each chromosome with geometric support in the CGR. The points from the chaos game representation are displayed one-dimensionally in sequential order. Red markers indicate the approximate location of centromeres.} 
    \label{Fig:figure16} 
\end{figure*}

In both cases, mitochondrial DNA is distinct from the other spectra. This is due partly to its length but mainly because its support differs significantly from that of chromosomes, as can be observed from the distributions. However, its spectrum provides new information about its tendency toward dimensions below references like the Sierpinski triangle, which has a fractal dimension of approximately 1.585. This is likely due, in part, to its bias toward regions corresponding to the A and C bases, considering that the G base accounts for only about $13\%$ of the genomic sequence, approximately half the frequency of the other bases.

\subsection{Cytogenetics of the CGR}\label{sec: Citogenética de la CGR}

Since the CGR respects the order of the sequence and the location of its points depends on previous states, it is possible to extend the one-dimensional CGR along the sequence to analyze potential changes. Figure \ref{Fig:figure15} shows the plot of each base's location in the GS versus the one-dimensional CGR obtained using the transformation \(x + \lfloor y(4^{6} - 1)\rfloor\). This transformation arranges the points similarly to Figure \ref{Fig:figure10}, extending them in a one-dimensional layout that follows the order of the genomic sequence.

Figure \ref{Fig:figure16} shows the corresponding plots for the GS of each chromosome and the approximate location of the centromere region, taken from databases in \citep{SG}.

This type of representation enables us to observe the characteristic banding patterns of chromosomes (Section \ref{sec: Citogenetica}), which facilitates, approximately, the identification of different sections of the GS within chromosomes or entire regions with high or low concentrations of certain bases. For instance, in chromosome 1, near the centromere (marked approximately in red), we can distinguish an almost CG-empty region, along with other smaller areas with distributions biased towards C and G bases. This could indicate potential coding and non-coding regions. However, the significance of this image from a multifractal perspective lies in showing that the sequence does not have a unique support structure across the GS of the chromosomes.

Base-by-base analysis is common in genomic studies, as it allows researchers to annotate each chromosome section in detail. This representation, combined with a dimensional measure, can aid in locating and characterizing regions of interest.

\subsection{MC and BGR Methods}

\begin{figure*}[t!] 
    \centering 
    \includegraphics[width=\linewidth]{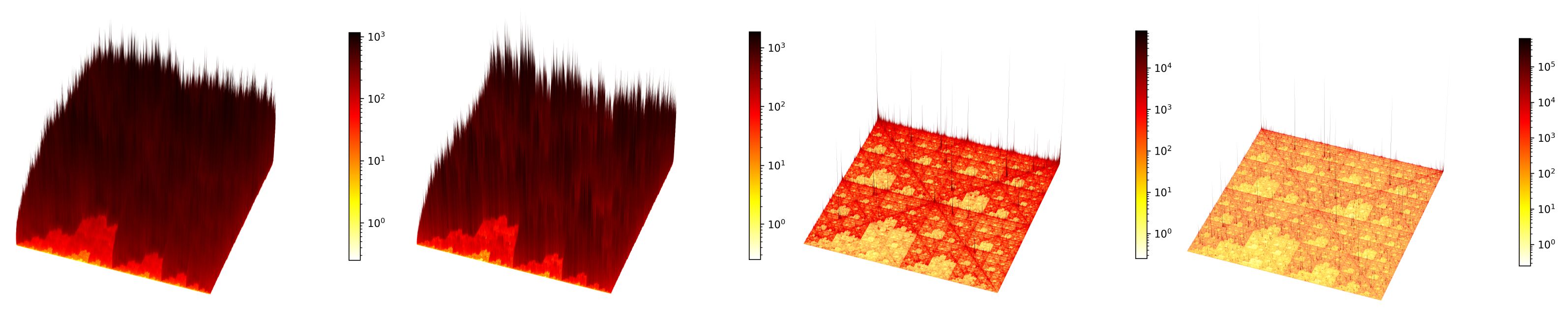}
    \caption{Point distributions generated from Markov Chains, with four different lengths forming $n$-nucleotides where $n={2, 3, 6, 12}$, from left to right.} \label{Fig:figure17} 
\end{figure*}

\begin{figure*}[t!] 
    \centering 
    \includegraphics[width=\linewidth]{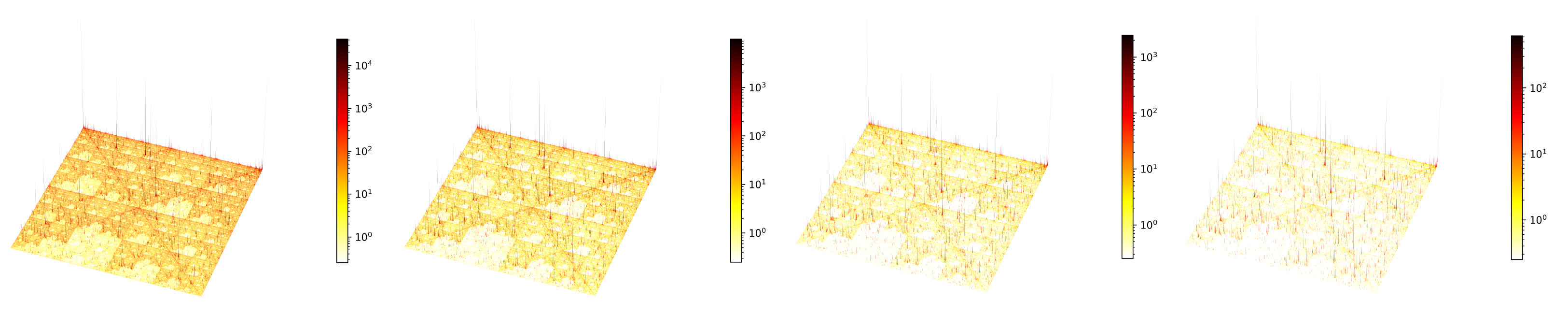}
    \caption{Point distributions generated from BGR, for $M={15, 63, 255, 1023}$, from left to right.} 
    \label{Fig:figure18}
\end{figure*}

Since the chromosome assembly exhibits behavior close to the chromosomal average, we can use it as a reference to evaluate the compression, encoding, and decoding methods of Markov Chains (MC) and Binary Genomic Representation (BGR) \label{Sec: GR}. Figures \ref{Fig:figure17} and \ref{Fig:figure18} display point distributions for sequences generated by these methods, using different length parameters. For MC, sequences of various lengths (n) were considered, creating probability matrices for dinucleotides, trinucleotides, hexanucleotides, and dodecanucleotides. Each matrix was used to generate sequences and CGRs matching the GS length of the assembly, i.e., $3,117,275,501$ bases. We observed notable differences in regions outside CG areas, especially for shorter chains; however, the fit improves for 6- and 12-base sequences.

The behavior is different for the BGR method, where, depending on the length, the number of points generated from the GS is much smaller than the original assembly. This reduction primarily impacts less represented areas while preserving structure in denser regions.

\begin{figure*}[t!] 
    \centering 
    \includegraphics[width=\linewidth]{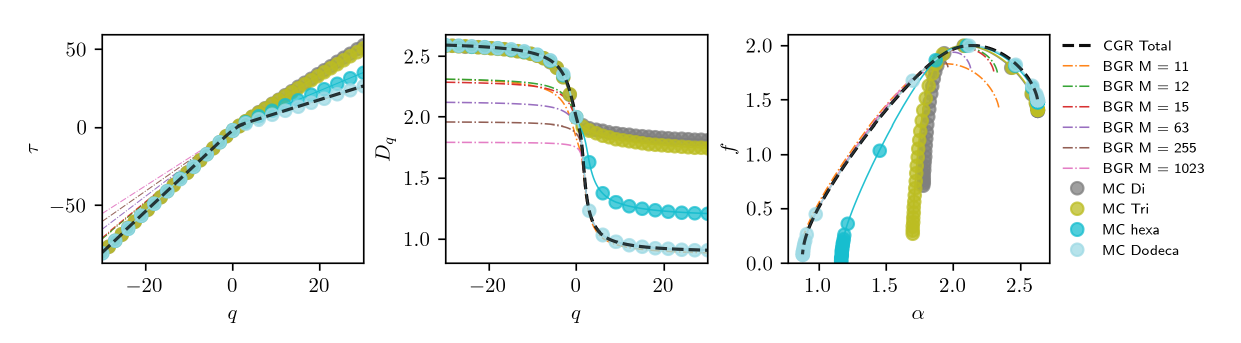}
    \caption{Multifractal analysis of BGR and MC methods compared to the complete CGR results.} 
    \label{Fig:figure19} 
\end{figure*}

The distribution biases observed with both methods are evident in their multifractal spectra compared to the complete CGR. Figure \ref{Fig:figure19} shows that, in low-incidence areas ($q < 0$) and CG regions of the CGR (see Figure \ref{Fig:figure4}), the BGR method loses precision as $M$ increases, impacting coding zones with higher compression rates. However, the fit in non-coding high-incidence regions remains nearly perfect. Importantly, BGR allows full decoding of the sequence, maintaining distribution characteristics in non-coding zones.

In MC, sequences tend to fit better in lower-populated areas because their length matches that of the assembly; however, shorter lengths can affect low-incidence areas. Precision significantly improves with longer sequences, reaching near-perfect fit with 12-base chains, suggesting that the optimal MC sequence length should equal the number of CGR partitions. Although MC aligns well with the multifractal spectrum of the assembly, it is not a coding or decoding method, as each value is generated randomly and does not reflect the chromosomal strip patterns. Nonetheless, the multifractal fit is remarkable, with an average error of approximately $2\%$ relative to the assembly in 12-base sequences.

\section{Discussion}

Although the GS exhibits multifractal characteristics that can be simulated using Markov Chains (MC), the fact that chromosomes retain fixed characteristics in typical humans means that the random nature of MCs is generally inconsistent, not accounting for genetic aspects that vary randomly among individuals. This randomness initially represented an advantage for encoding and decoding methods. However, it seems evident that a combination of the two methods could improve the fit to fixed multifractal and structural characteristics in chromosomes, while also incorporating the random traits specific to humans. This approach presents a considerable challenge, as it would require sequences from different individuals for comparative analysis.

Regarding the structural and cytogenetic aspects of the GS, an analysis considering the variations or fluctuations along each chromosome is relevant to relate chromosomal parts to the complete assembly, as well as to identify shared patterns or specific ones that could be linked to specific biological functions based on multifractal properties. Given the observed bias in chromosomes 9 and Y, they seem like a good starting point for this type of analysis. Additionally, due to its unique behavior, it might be more appropriate to analyze mitochondrial DNA separately.

\section{Conclusions}

The main conclusions from the analysis of the complete genomic sequence and individual chromosomes are as follows: the multifractal structure is preserved throughout the sequence, despite each chromosome exhibiting a unique distribution. Chromosomes maintain an approximately constant fractal support, but the variations in their distributions may reflect chromosome-specific functions, even if these functions are not yet fully understood. Additionally, the CGRs for both the full assembly and individual chromosomes effectively differentiate between repetitive and non-coding regions, CpG islands, and coding sections across all CGR regions.

The Markov model successfully simulates the multifractal structure of genomic sequences when the chain length matches the number of divisions in the CGR's containing square. However, it fails to replicate the fixed stripe patterns observed in chromosomes. Consequently, the binary genomic representation is a more suitable method for compacting, encoding, and decoding genomic information and for approximating the multifractal structure of chromosomes.

Mitochondrial DNA, in contrast, exhibits entirely distinct characteristics in both composition and multifractal structure. As a shorter sequence, it shows more pronounced fractal support in regions rich in A, C, and T bases, expanding empty regions in the CGR. This results in narrower ranges for multifractal measures — including $\tau$ exponents, singularity, generalized dimension, and spectrum — compared to those observed in chromosomes.


\section*{Acknowledgments}
\noindent Y.A.A.-B. received funding for his doctoral studies from CONAHCyT under its national program of scholarships for graduate studies (CONAHCyT, 2021-2). M.A.Q.-J. and  A.M.E.-R. thankfully acknowledges financial support by CONAHCyT under the Project CF-2023-I-1496. Also, M.A.Q.-J. would like to thank the support from DGAPA-UNAM under the Project UNAM-PAPIIT IA103325. A.M.E.-R. thanks the support from UAM research grant 2024-CPIR-0.


\bibliographystyle{ieeetr}
\bibliography{Manuscript}

\end{document}